\documentclass[aps,nofootinbib,twocolumn,preprintnumbers,superscriptaddress,nobibnotes,floatfix,longbibliography]{revtex4-1}

\pdfoutput=1
\usepackage{amsmath,amsfonts,amssymb,mathrsfs}
\usepackage{hyperref,cleveref}
\usepackage[utf8]{inputenc}
\usepackage{esint}
\usepackage{graphicx}
\usepackage{color}
\usepackage{ragged2e}
\usepackage{tikz}
\usepackage[loose]{units}

\begin{document}

\title{Dynamical Dark Energy Emerges from Massive Gravity}


\author{Juri Smirnov\\
\href{mailto:juri.smirnov@liverpool.ac.uk}{juri.smirnov@liverpool.ac.uk} \,
\href{https://orcid.org/0000-0002-3082-0929}{ORCID: 0000-0002-3082-0929}
}
\affiliation{Department of Mathematical Sciences, University of Liverpool, Liverpool, L69 7ZL, United Kingdom}

\begin{abstract}
In this work, we demonstrate that a dynamical dark energy component predicted by massive gravity gives rise to a distinctive evolution of the equation of state. This scenario is favoured over the standard $\Lambda$CDM model when confronted with the latest combined datasets from the Dark Energy Spectroscopic Instrument (DESI), the cosmic microwave background (CMB), and supernova observations. The model stands out as a rare example of a healthy, self-consistent theory that accommodates phantom dark energy while maintaining a technically natural, small asymptotic cosmological constant. Our analysis indicates a preferred graviton mass of approximately $4.0 \times 10^{-33} \text{eV}$, suggesting the emergence of a new cosmological length scale. This leads to a maximal deviation of the equation of state around $z \sim 3$, a prediction that will be robustly tested by upcoming, deeper surveys of baryon acoustic oscillations.
\end{abstract}

\preprint{LTH-1404}
\maketitle

\section{Introduction}
The discovery of accelerated expansion showed that the universe is more complex than previously thought~\cite{SupernovaCosmologyProject:1998vns}.  With increasing data, we wonder if dark energy is more complex than just a cosmological constant.  If so, it could be a dynamical quantity, with a time-varying equation of state~\cite{Wetterich:1987fm,Dymnikova:2000gnk,Copeland:2006wr,Bhattacharya:2024kxp}. Recent findings by the Dark Energy Spectroscopic Instrument (DESI) suggest that this may indeed be the case~\cite{DESI:2025fii,Scherer:2025esj}.


In this Letter, we consider this question in the theoretical framework of a conceptually minimal extension of general relativity (GR) including one massive spin-2 field~\cite{Pauli:1939xp,Fierz:1939ix,vanDam:1970vg,Zakharov:1970cc,Vainshtein:1972sx,Boulware:1973my,deRham:2010ik,deRham:2010kj,Hassan:2011vm,Hassan:2011hr,Hassan:2011tf,Hassan:2011ea,Hassan:2011zd,deRham:2014zqa,deRham:2014naa,deRham:2014fha}. As discussed in Refs.~\cite{deRham:2010ik,deRham:2010kj,Hassan:2011zd}, the unique, non-pathological, effective theory that has a massive graviton in its spectrum relies on an intertwining of metric fields, with its minimal representation being bigravity featuring two metric fields g and f~\cite{Hassan:2011zd}. 

An interesting property of this theory is that it can exhibit a large degree of symmetry:
\begin{align}
\text{Diff}(g) \otimes \text{Diff}(f),
\end{align}
where $\text{Diff}$ are the diffeomorphisms acting on the metric fields g and f. As we discuss below, the introduction of a mixing potential between the two fields leaves only the diagonal subgroup diffeomorphism invariant. The object transforming as a singlet is given by the determinant of the field contraction $ \kappa = g^{-1} \, f$. The Lagrangian terms depending on $\kappa$ determine the mass scale of the massive spin-2 field propagating in the theory, and, as we will demonstrate, also set the scale of the induced dark energy, leading to an accelerated expansion. Thus, in the limit where the terms that contain $\kappa$ vanish, we have an enhanced total symmetry in our theory and also vanishing dark energy, making the small numerical value with respect to the Planck scale technically natural (see also Refs.~\cite{Hinterbichler:2011tt,deRham:2014zqa}).
\begin{figure}[h]  
        \centering
    \includegraphics[width=0.9\columnwidth]{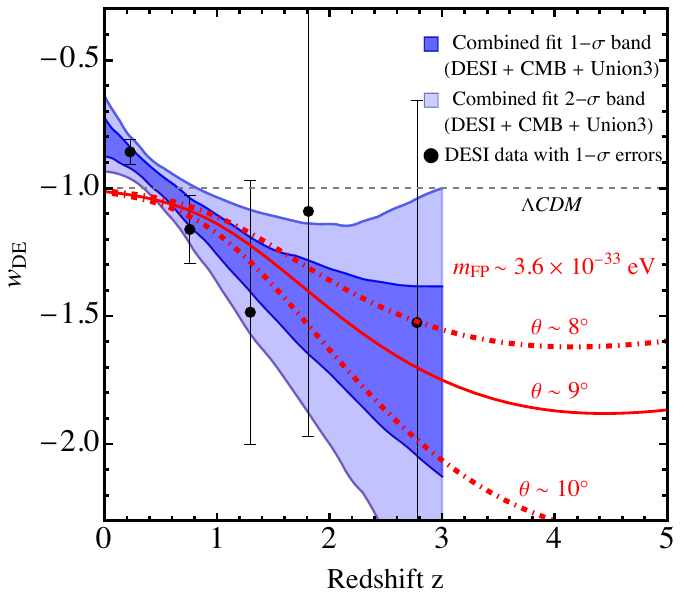}
        \caption{The dark energy equation of state parameter in bigravity (red lines). Graviton mass obtained from the fit to the data obtained in the Gaussian Process Regression (GPR) analysis of the DESI, CMB, and Union3 datasets~\cite{DESI:2025fii}.
        }
        \label{fig:DESIfit}
\end{figure}

It is interesting that in the context of bigravity, the cosmological solution unavoidably features a dynamical time-evolving dark energy component $\rho_{\rm DE}$. In cosmologies that exhibit late time deSitter behaviour, its equation of state is always phantom, i.e., $w_{\rm DE} < -1$~\cite{Singh:2003vx}, and only asymptotically approaches $w = -1$ at early and late times. Thus, the theory provides a healthy realisation of a phantom dark energy scenario, which is a highly non-trivial condition~\cite{Notari:2024rti}. The maximal deviation from this asymptotic value appears at length scales related to the mass of the spin-2 field in the theory, thus, the theory has an additional explicit length scale in addition to the Planck scale~\cite{Luben:2019yyx}.

Figure~\ref{fig:DESIfit} previews our results, for which the methods are detailed below. 
We observe that the reconstructed evolution of the equation of state parameter, $w_{DE}$, from  DESI~\cite{DESI:2025fii}, CMB~\cite{Planck:2018nkj}, and Union3~\cite{Rubin:2023ovl} supernova datasets is very well reproduced in bimetric gravity at $z>0.5$ and points to a graviton mass scale of $m_{\rm FP} \sim 4 \cdot 10^{-33} \, \text{eV}$. At small $z$, the theory shows deviations from the DESI results comparable to the standard $\Lambda \text{CDM}$ cosmology. Intriguingly, the evidence for $w_{\rm DE} > -1$ at low $z$ is weaker in the new DESI data set, see Fig. 3 in Ref.~\cite{DESI:2025wyn}. The additional length scale associated with the graviton mass predicts a specific evolution of the dark energy equation of state parameter. Importantly, this can be verified with deeper observations at redshifts, $z \geq 3$.


 \section{Massive Gravity and the Bigravity Framework}
Currently, the only consistent way of formulating a theory with a massive spin-2 field is the introduction of a reference metric $f$, with a uniquely defined interaction potential $V(g,f)$~\cite{deRham:2010ik,Hassan:2011zd}. Bigravity generalises this ansatz by making $f$ a dynamical field, resulting in a symmetric system with two interacting metric tensors, $g_{\mu\nu}$ and $f_{\mu\nu}$~\cite{Hassan:2011zd,Hassan:2011tf}. The theory is governed by the action:
\begin{align}
S = \frac{M_g^2}{2} \left( \int d^4x \sqrt{-g} R(g) + \alpha^2 \int d^4x \sqrt{-f} R(f) \right)  \\ \nonumber - M_g^2 \int d^4x \sqrt{-g} V(g, f) + \int d^4x \sqrt{-g} \mathcal{L}_m(g, \Phi),
\end{align}
where $M_g$ is the Planck mass corresponding to the $g$ metric, and $\alpha$ the mass scale ratio for the $f$ metric expressed as $M_f = \alpha M_g$, $R(g)$, and $R(f)$ are the Ricci scalars of the respective metrics, and $V(g,f)$ is the interaction potential ensuring ghost-free dynamics. Finally, $\mathcal{L}_m(g, \Phi)$ is the standard model (SM) Lagrangian, which can be consistently incorporated through minimal coupling to a single metric~\cite{Hassan:2012wr,deRham:2014naa}.

The interaction potential takes the form:
\begin{equation}
V(g, f) = \sum_{n=0}^{4} \beta_n e_n(\sqrt{g^{-1} f}) = \sum_{n=0}^{4} \beta_n e_n(\sqrt{\kappa}),
\end{equation}
where $e_n$ are elementary symmetric polynomials and $\beta_n$ are free parameters that determine the strength of the interaction, and have mass dimension two in our paramterisation.
Varying the action with respect to both metrics leads to two sets of Einstein equations:
\begin{align}
G^g_{\mu\nu} + V^g_{\mu\nu} &= \frac{1}{M_g^2} T_{\mu\nu}, \\
G^f_{\mu\nu} + V^f_{\mu\nu} &= 0,\\
\nabla^\mu V^g_{\mu\nu} &= 0 \label{eq:bianchi},
\end{align}
where $G^g_{\mu\nu}$ and $G^f_{\mu\nu}$ are the Einstein tensors for $g_{\mu\nu}$ and $f_{\mu\nu}$ respectively, while $V^g_{\mu\nu}$ and $V^f_{\mu\nu}$ arise from the variation of the potential, see for example Ref.~\cite{Platscher:2018voh} for their explicit form. The last expression, Eq.~\ref{eq:bianchi}, is the Bianchi constraint and follows from covariant energy-momentum conservation.  

Before discussing the solutions to the above equations of motion in the cosmological context, it is instructive to consider a static, spherically symmetric system with a Schwarzschild-type solution. As discussed in Refs.~\cite{Babichev:2013pfa,Platscher:2016adw,Platscher:2018voh}, the gravitational potential in this case takes the form
\begin{align}\label{eq:gravitational-potential}
    \phi(r) =
    - \frac{1}{M_g^2} \frac{1}{r} \begin{cases}
        1  & r \ll r_{\rm V}\,, \\
           \cos^2(\theta) + \frac{4 \sin^2(\theta)}{3}\, e^{ -m_\text{FP} r}   & r \gg r_{\rm V} \,,
    \end{cases}
\end{align}
where $m_\text{FP}$ is the mass of the massive spin-2 mode, given by
\begin{align}
    m_{\rm FP}^2  = \frac{1}{\sin^{2}(\theta)} \left(  \beta_1 c + 2 \beta_2 c^2 + \beta_3 c^3 \right)\,.
\end{align}
The $\theta$ angle determines the degree of mixing between the massive and the massless modes~\cite{Max:2017kdc,Max:2017flc}. The mixing is related to the ratio of the mass scales for the $f$ and $g$ metrics by $\sin^2(\theta) = (1 + \bar{\alpha}^{-2})^{-1}$, and $\cos^2(\theta) = (1 + \bar{\alpha}^2)^{-1}$. In a general solution to the bigravity equations of motion, in the absence of sources, theoretical consistency requires that $f$ and $g$ must be proportional i.e. $f_{\mu \nu}  = c^2 \ g_{\mu \nu}$, and the constant $c$ enters the graviton mixing parameters as $\bar{\alpha} = \alpha c$.

The Vainshtein radius, $r_V$, marks the scale below which the general relativity potential is restored since the non-linearities of the fields become so strong that the massive mode does not propagate~\cite{Babichev:2013pfa,Babichev:2013usa}
\begin{align}
r_{\rm V} = \left(\frac{r_{\rm S}}{m_\mathrm{FP}^2}\right)^{1/3}\,,
\end{align}
where the Schwarzschild radius for a source of mass $M$ is $r_{\rm S} =  M/(4 \pi \, M_g^2)$.

The Vainshtein screening mechanism is crucial for the phenomenological viability of the model. However, not all combinations of the $\beta_i$ parameters lead to successful screening. Introducing the variables
\begin{align}
   & A = \frac{\beta_2 c^2 + \beta_3 c^3}{ m_{\rm FP}^2 \sin^2(\theta)}\,, \\
   &  B = \frac{\beta_3 c^3}{m_{\rm FP}^2 \sin^2(\theta)} \,,
\end{align}
Ref.~\cite{Hogas:2021fmr,Hogas:2021lns} shows the viable parameter space for a successful screening around localised sources (Fig. 8).

A crucial observation can be made by considering the length scale, $l_{\rm EFT} = (M_{g} m_{\rm FP}^2)^{-1/3}$, below which the bimetric effective description breaks down (see Refs.~\cite{deRham:2010ik, Akrami:2015qga} for a derivation). We can rewrite the Vainshtein radius for a gravitating object as $r_V = l_{\rm EFT} \,   (M/(4 \pi M_g))^{1/3}$. We see that for masses above the Planck scale $r_V \gg l_{\rm EFT}$, and thus GR is restored before the length scales at which the bigravity EFT breaks down. Therefore, from a theoretical point of view, bigravity fully describes classical gravitating systems and would require a UV completion that involves an understanding of quantum gravity only when dealing with distances close to the Planck length, similar to GR. 

\section{Cosmological Solutions}
We now turn to the explicit solution for the bigravity equations of motion in a homogeneous and isotropic universe~\cite{DAmico:2011eto,vonStrauss:2011mq,Comelli:2011zm,Akrami:2012vf,Volkov:2011an,Konnig:2013gxa,DeFelice:2014nja}. The ansatz for both metrics takes the FLRW form:
\begin{align}
ds^2_g &= a^2(\eta) \left(-d\eta^2 +  d\mathbf{x}^2\right), \\
ds^2_f &= b^2(\eta) \left( -(1 + \mu(\eta))^2 d\eta^2 + d\mathbf{x}^2 \right).
\end{align}
Here, $\eta$ is the conformal time, $a(\eta)$ and $b(\eta)$ are scale factors, and $1 + \mu(\eta)$ represents the lapse function for the $f$-metric. Here, $\mu(\eta)$ can be identified with the Stückelberg field that signals the transition from the strong to weak coupling regime for the massive mode~\cite{Luben:2019yyx}.

If we write the conformal Hubble parameters as $H = \dot{a}/a$ and $\tilde{H} = \dot{b}/b$, and define $y = b/a$, we find that the Bianchi constraint, Eq.~\ref{eq:bianchi}, leads to
\begin{align}
    \left(H (1+ \mu) - \tilde{H} y \right) \left(\beta_1 + 2 \beta_2 y + \beta_3 y^2 \right)= 0\,,
\end{align} 
which has two branches of solutions. Setting the second bracket to zero corresponds to the algebraic branch, which is pathological~\cite{Comelli:2012db}, so only the $H (1 + \mu) = \tilde{H} y$ constraint leads to physical solutions on the so-called dynamical branch~\cite{Akrami:2015qga}. This can be written as
\begin{align}
    \mu = \frac{\dot{y}}{H y} = \frac{y'}{y}\,,
\end{align}
where we introduce the derivative w.r.t. $e$-folds: $' = d/d \ln (a)$. Note that the conformal and physical Hubble parameters are related via $H = H_{\rm conf}  = a H_{\rm phys}$.

The system of Friedmann equations for the two metrics can be written as follows:
\begin{align}
3 M_g^2 H^2 &= a^2 (\rho_m + \rho_{\text{DE}}(y)),  \label{eq:FR1} \\
3 H^2 &= a^2 \Lambda(y).
\label{eq:FR2}
\end{align}
The first equation contains the source term that drives the Hubble evolution based on the matter-energy density, $\rho_m$, and the time-evolving dark energy component of the universe, $\rho_{\text{DE}}$. The second equation provides an additional constraint based on the dynamics of the second metric, 
where 
\begin{align}
 \Lambda(y) = \frac{1}{\alpha^2} (\beta_1 y^{-1} + 3\beta_2 + 3\beta_3 y + \beta_4 y^2)\, 
\end{align}
 will asymptotically approach the cosmological constant in the system. The effective dark energy density is explicitly given by:
\begin{equation}
\label{eq:DDE}
\rho_{\text{DE}}(y)/M_g^2 = \beta_0 + 3\beta_1 y + 3\beta_2 y^2 + \beta_3 y^3.
\end{equation}
We observe that the model exhibits a dynamical dark energy component that evolves with redshift, deviating from the standard $\Lambda$CDM behavior, which we will discuss in the next section.

However, let us first discuss the general behaviour of the cosmological solution in massive gravity.  Combining the Eqs.~\ref{eq:FR1}, and \ref{eq:FR2} leads to a fourth-degree polynomial in $y$ with up to four real-valued roots. As discussed in Refs.~\cite{Koennig:2013fdo, Konnig:2015lfa} only one of the roots corresponds to a healthy physical solution in which the scale ratio evolves from $y \rightarrow 0$ in the early universe to a constant value $y \rightarrow c$ at late times, where the universe asymptotes to its vacuum solution, the so-called finite branch. 

Now, let us define time-evolving quantities that approach constant vacuum values in the asymptotic future. The quantity $\rho_{\rm DE}(y)$ and $\Lambda(y)$ evolve towards $ \rho_{\rm DE}(c)= \Lambda(c) M_g^2$ for $\eta \rightarrow + \infty$, which can be seen as the asymptotic vacuum energy of the system.
Additionally, we introduce the quantities:
\begin{align}
\sin^2(\theta)(y) = \frac{1}{\left( 1 + (\alpha^2 y^2)^{-1}\right)}\,,
\end{align}
which can be seen as the mode mixing between the massive and massless modes, and:
\begin{align}
    m_{\rm FP}^2(y) = \frac{1}{\sin^{2}(\theta)(y)} \left(  \beta_1 y + 2 \beta_2 y^2 + \beta_3 y^3 \right)\, ,
\end{align}
which is the mass of the non-linear massive mode on the FRW background~\cite{Luben:2019yyx}. We observe that in the limit of $\alpha \rightarrow 0$, the mass $m_{\rm FP} \rightarrow \infty$ and the massive dynamics fully decouples, physically restoring GR. The time evolution of $m_{\rm FP}(y)$ and $\sin^2(\theta)(y)$ is such that at late times $m_{\rm FP}(y) \rightarrow m_{\rm FP}(c) = m_{\rm FP}$, and $\sin^2(\theta)(y) \rightarrow \sin^2(\theta)(c)$, which are the vacuum graviton mass, and mixing given in Eq.~\ref{eq:gravitational-potential}, respectively.

Another important quantity is the Stückelberg field $\mu$. Taking the $e$-fold derivative of the polynomial determining $y$ and using the conservation equation
\begin{align}
    \dot{\rho}_m + 3 H \left( 1 + w_m \right) \rho_m = 0 \, ,
\end{align}
leads to its evolution Eq.~\cite{Luben:2019yyx}
\begin{align}
    \mu(y) = \frac{y'}{y} = \frac{(1 + w_m) \rho_m/M_g^2}{m_{\rm FP}^2(y) - 2 H^2}\,.
\end{align}
Since on the finite branch this quantity must not change its sign, this implies a physical constraint $m_{\rm FP}^2(y) > 2 H^2$, the so-called Higuchi bound~\cite{Fasiello:2013woa,Konnig:2015lfa}. For the asymptotic values of the graviton mass and the vacuum energy, this condition leads to
\begin{align}
m_{\rm FP}^2 > \frac{2}{3} \Lambda(c)\,, 
\end{align}
which means that the graviton mass has a physical lower bound, set by the scale of the asymptotic cosmological constant in our universe. 

The time evolution of the Stückelberg field, as discussed in Ref.~\cite{Luben:2019yyx}, determines the transition from the regime where the massive mode is strongly coupled, and the matter-energy density $\rho_m$ dominates the dynamics and $\mu \sim 3 (1 + w_m)$, and the regime where $\rho_{\rm DE}(y)$ starts to become dominant, $\mu \rightarrow 0$, and the universe approaches a deSitter phase for $\Lambda(c) >0$. Crucially, it is the mass scale of $m_{\rm FP}$ that determines the value of the Hubble rate at which this important transition takes place, and the strongest variations in the dark-energy equation of state occur, as we discuss shortly.
 
Note that in the case where all the $\beta_i$ parameters in the theory are set to zero, the system exhibits the larger diffeomorphism symmetry $\text{Diff}(g) \otimes \text{Diff}(f)$ acting on the $g$ and $f$ metrics and, at the same time, $\rho_{\rm DE}(c)/M_g^2 = \Lambda(c) =0$ vanishes identically. Hence, a small value for the evolving and asymptotic dark energy is technically natural in massive gravity.

\section{Dynamical Dark Energy in Bigravity}
In bimetric theory, the evolution Eq.~\ref{eq:FR1} features a time-evolving source $\rho_{\rm DE}(y)$, given by Eq.~\ref{eq:DDE}, that represents a dynamical dark-energy component of the universe. In the asymptotic future $\rho_{\rm DE}(y) \rightarrow \Lambda(c) M_g^2$, which indicates a deSitter final stage for $\Lambda(c) >0$.  As derived in~\cite{Luben:2019yyx}, the equation of state for this parameter $\rho_{\rm DE}(y)$ is given by:
\begin{align}
    w_{\rm DE} = -1 - \sin^2(\theta)(y)  \frac{m_{\rm FP}^2(y) M_g^2}{\rho_{\rm DE}(y)} \, \mu(y)\,.
    \label{eq:EOS}
\end{align}

\begin{figure}[h]  
        \centering
        \includegraphics[width=0.9\columnwidth]{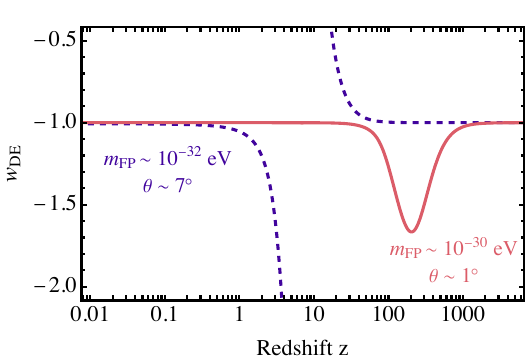}
        \caption{The dashed line shows the behaviour of $w_{DE}$ if $\rho_{DE}(y)$ has a sign flip around $z \sim 10$, while the solid line shows the EOS-evolution for a spacetime with larger graviton mass and $\rho_{DE}(y) >0$.}
        \label{fig:SignFlip}
\end{figure}

Figure~\ref{fig:SignFlip} shows two evolution scenarios for the dark energy equation of state (EOS). We observe that the sign of the dynamical dark energy component determines whether the EOS parameter increases above $w = -1$ (quintessence regime) or whether it decreases below $w = -1$ (phantom regime).
Note that the system can feature a transition from the quintessence to the phantom regime, however, this requires $\rho_{\rm DE}(y)$ to switch sign, which leads to a discontinuous evolution of the EOS-parameter. 
In the case that the universe is an asymptotic deSitter spacetime with $\Omega_\Lambda >0$ the EOS is always phantom at late times. 

As can be seen from Eq.~\ref{eq:EOS}, the strongest deviations from the constant equation of state behaviour are expected when the Stückelberg field, $\mu(y)$, undergoes the regime change between the state where the massive mode is strongly coupled and the regime where it freely propagates~\cite{Luben:2019yyx}. This process is very similar to the radial evolution of the Stückelberg field in a Schwarzschild background that takes place around the Vainshtein radius, determined by the mass of the source and the graviton wave-length~\cite{Babichev:2013usa}. We will investigate this phenomenon numerically in the next section. 

\section{Fit to Observations}
\label{sec:fittoobservations}
An initial analysis of the data in Ref.~\cite{DESI:2025fii} can be made in a somewhat restricted model, where only the parameters $\beta_0, \beta_1, \beta_4$ are non-zero~\cite{Luben:2020xll}. This fit yields a graviton mass scale of the order of $(5.0 \pm 0.5)\times 10^{-33} \rm{eV}$ and exhibits a phantom dark energy EOS. However, as discussed in Ref.~\cite{Hogas:2021fmr} this model with the screening parameters $A = B = 0$ does not lead to Vainshtein screening around localised sources. Thus, this simplifying assumption needs to be relaxed, and a full fit to five model parameters is required. 

For the fit to the DESI data at small redshift values, we rewrite Eq.~\ref{eq:EOS} in the following convenient form:
\begin{align}
   w_{\rm DE} =  -1 -  \frac{\sin^2(\theta)(y)\, \Omega_{\mathrm{source}} }{ {\Omega_{\mathrm{DE}}(y)}\left(1 - \frac{2 H^2}{ m_{\mathrm{FP}}^2(y)}\right)}\,,
\end{align}
where the source term is dominated by non-relativistic matter $\Omega_{\mathrm{source}} \approx \Omega_{m0}\,(1+z)^3$. Combining Eqs.~\ref{eq:FR1} and~\ref{eq:FR2} we obtain a fourth-degree polynomial for $y$ as a function of the source density that scales with $z$ (see App.\ref{app:parametrisation} for its explicit parametrisation). Solving it numerically, and choosing the physical branch, on which $ 0 < y/c \leq 1$, we obtain the evolution of the dark energy EOS. The quantities $\Omega_{\rm DE}(y)$, $m_{\rm FP}(y)$, and $H(y)$ are expressed in terms of the parameters $\Omega_\Lambda$, $H_0$, $\Omega_{m0}$, $m_{\rm FP}$, $\theta$, $A$, and $B$, (see App. \ref{app:parametrisation}).
This paramtrisation is convenient for the fit to the experimental data. At the same time, the existence of the local Vainshtein screening can be ensured by requiring that the parameters $A$ and $B$ are within their allowed values given a fixed graviton mixing angle $\theta$, see Ref.~\cite{Hogas:2021fmr}.

In our analysis, we focus on the preferred region for the equation of state parameter evolution based on Fig. 9 in Ref~\cite{DESI:2025fii}. Here, Gaussian Process Regression has been applied to the combined data sets of DESI, Union3, and the CMB data, thus covering data based on baryon-acoustic oscillations (BAO), supernovae, and inhomogeneities in the early universe, respectively. The analysis represents a rather robust estimate for the preferred values of the equation of state parameter, without explicit dependence on the choice of redshift binning. In Fig.~\ref{fig:DESIfit}, we show the evolution of the equation of state $w_{\rm DE}$ in bigravity with three example values of mixing parameters at a graviton mass value chosen from the best-fit region to the combined observational data of DESI, CMB, and Union3. 

\begin{figure}[h]  
        \centering
        \includegraphics[width=0.9\columnwidth]{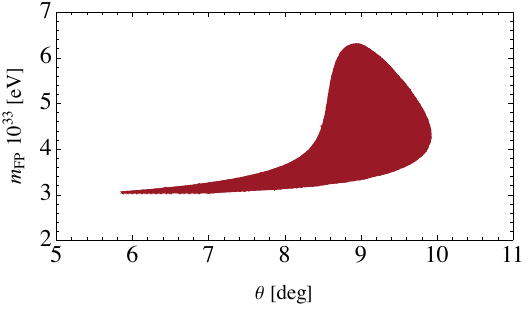}
        \caption{The best-fit contour of the graviton mass and mixing parameters at the one sigma level. The remaining bimetric parameters are marginalised over.} 
        \label{fig:BestFit}
\end{figure}

Figure~\ref{fig:BestFit} shows the one and two sigma contours for the graviton mass and mixing parameter that best fit the data. 
The best fit intervals of the remaining bimetric parameters are $A = - 0.49 \pm 0.03$, $B = 2.80 \pm 0.03$, and the asymptotic dark energy value $\Omega_\Lambda = \Lambda(c)M_g^2/\rho_c \sim 0.70\pm 0.12$. We adopt a continuous version of the $\chi^2$ test, computing the ratio of integrals over the quadratic deviations relative to the mean squared errors between the data and the two theoretical models: $\Lambda$CDM and bigravity.  The obtained best-fit parameter choice in bigravity leads to a $75\%$ improvement of the fit to the GPR data over the $\Lambda \text{CDM}$ model.  We also observe that a fit to the binned DESI data alone does not significantly improve the fit over a constant equation of state scenario.  The best-fit parameters in bigravity are fully compatible with observations of galaxies and galaxy clusters~\cite{Platscher:2018voh}, observational~\cite{Hogas:2021lns}, and theoretical constraints~\cite{Hogas:2021fmr}.

The graviton mass preferred by the GPR data indicates a new physical scale that leads to the maximal variation of the dark energy EOS parameter around the redshift $z \gtrsim
 3$. At small redshifts, the equation of state approaches $w = -1$, thus naturally avoiding a big-rip scenario. Note that in a universe with one massive graviton mode, and an asymptotic value of $\Lambda(c)>0$ the dark energy equation of state is always $w_{\rm DE} < -1$, thus it does not match the one-sigma region of the low redshift bin of the current DESI data release in the binned analysis of Ref.~\cite{DESI:2025fii}. However, intriguingly, a recent unbinned analysis sees weaker evidence for $w_{\rm DE} > -1$ at low $z$ in the DESI Data Release 2 compared to DESI Data Release 1, see Fig. 3 of Ref~\cite{DESI:2025wyn}. It remains to be seen whether $w_{\rm DE}$ crossing $w = -1$ will be confirmed with more BAO data. 

In addition to the fit to the GPR data, we adopted the method detailed in Ref.~\cite{Luben:2019yyx} to check whether the predicted cosmological evolution is in agreement with CMB and supernova observations in a simplified framework. To this end, the deceleration parameter $q$, the angular diameter distance normalised to the Hubble horizon at CMB decoupling,  the CMB principal multipole number, and the energy density of baryons at CMB decoupling are computed in the full bigravity framework and compared to observations.

We find that the best-fit point in bigravity is preferred over the $\Lambda \text{CDM}$ model by $\Delta \chi^2/\text{dof} \sim 4$. It is intriguing to observe that this improvement results from a Hubble rate that is in agreement with CMB observations, but pulled to higher values at late times. In this case $H_0 \sim 70.2 \pm 0.7 \, \text{km}/\text{s}/\text{Mpc}$, compared to $H_0 \sim 68.0 \pm 0.6 \, \text{km}/\text{s}/\text{Mpc}$ preferred in $\Lambda$CDM. This can reduce the tension with local observations, which favour values closer to $H_0 \sim 73 \pm 1.7\, \text{km}/\text{s}/\text{Mpc}$. 


As recently discussed in Ref.~\cite{Dwivedi:2024okk}, the best fit value of the Hubble rate obtained in a bigravity fit to CMB and supernova data is $H_0 \sim 71.0 \pm 0.9 \, \text{km}/\text{s}/\text{Mpc}$, and is in agreement with local measurements. The best fitting bigravity parameters obtained in Ref.~\cite{Dwivedi:2024okk} are compatible with the parameters needed to explain the dynamical EOS in our scenario, especially since the graviton mass required only has to satisfy an upper bound that is orders of magnitude larger than the value we obtain in our analysis. An intriguing question remains of whether the crossing from the phantom regime to the quintessence regime can be confirmed with more observations, since it is harder to reconcile this dynamics with the Hubble tension~\cite{Colgain:2025nzf}.

A global analysis that determines whether the same model can explain the dynamical dark energy suggested by Ref.~\cite{DESI:2025fii} can solve the Hubble tension is certainly intriguing, but beyond our current scope. Furthermore, extending the framework to more than one massive spin-2 field~\cite{Flinckman:2024zpb}, might allow further improving the fit to the data, for example by accounting for the transition of $w_{\rm DE}$ from values below $w  = -1$ to values above $w =-1$ at low redshifts. Those additional extensions beyond the minimal scenario are left to forthcoming work. 

\vspace{-0.35cm}
\section{Conclusions}
\vspace{-0.35cm}
We demonstrate that, given the new DESI data release, bigravity is a viable and compelling alternative to $\Lambda$CDM with a technically natural small scale for the cosmological constant. Furthermore, bigravity provides a natural explanation for the evolving dark energy equation of state and is even favoured over $\Lambda$CDM by combined observations of the CMB, Union3, and DESI data. 
The preferred graviton mass can potentially reduce the observed tension between local measurements of the Hubble rate and its global determination from the CMB. The predicted graviton mass of $\sim \mathcal{O}(1) 10^{-33} \, \text{eV}$ points towards a new length scale in the universe that leads to maximal variation of the dark energy equation of state around $z \gtrsim 3$. Future DESI data releases and additional BAO analyses will provide a conclusive test of this scenario in the near future.

\vspace{-0.55cm}
\section*{Acknowledgments}
\vspace{-0.45cm}
I thank John Beacom, Spencer Griffith, Marcus Högås, and Jack Shergold for helpful discussions and comments on the manuscript. I also thank Roksolana Kundik for being a continuous inspiration and an amazing source of creativity. As a Simons Visiting Fellow (Award Number:1023171),  I thank the IIP in Natal for the hospitality during the final stages of this project. I acknowledge support from the UK Research and Innovation Future Leader Fellowship~MR/Y018656/1.

\newpage
\onecolumngrid
\appendix

\section{Bimetric Parametrisation}
\label{app:parametrisation}

In this appendix, we provide the explicit form of the bimetric parametrisation used in our analysis of the cosmological dynamics, see Ref.~\cite{Hogas:2021fmr}. The modified Friedmann equations, governing the evolution of the normalised scale factor ratio $\tilde{y}= b/(a c)$ in the bimetric framework, can be rewritten to obtain a quartic polynomial:
\begin{align}
& -(1 + 2A + B) \, \Omega_{\rm FP}  \cos^2(\theta) 
+  B \, \Omega_{\rm FP}  \, \tilde{y}^4 \sin^2(\theta) 
- \tilde{y}^3 \left( \Omega_{\Lambda} + (-1 + A - B) \, \Omega_{\rm FP}  \cos^2(\theta) + 3(A + B) \sin^2(\theta) \right) \\ \nonumber  
& + 3\, \Omega_{\rm FP}  \tilde{y}^2 \left( -B \cos^2(\theta) + (1 + 2A + B) \sin^2(\theta) \right) 
+ \tilde{y} \left( \Omega_{\Lambda} + (1 + A + B) \, \Omega_{\rm FP} \left((A + B) \cos^2(\theta) - \sin^2(\theta)\right) + \Omega_{\mathrm{source}}[z] \right) = 0\,,
\end{align}
where $\Omega_{\rm FP}  = m_{\rm FP}^2/(3 H_0^2)$. The normalisation of $\tilde{y}  = y/c$ has been introduced, with $c$ being the proportionality constant of the metrics in the asymptotic vacuum solution, to ensure that $\tilde{y}$ is bounded by one on the finite branch. The term driving the evolution is $\Omega_{\mathrm{source}}[z]$, which contains the cosmological source fluids and can be written as:
\begin{align}
    \Omega_{\mathrm{source}}[z] = \Omega_{\rm m0}(1+z)^3  + \Omega_{\rm r0}(1+z)^4\,, 
\end{align}
for a system that contains non-relativistic matter and radiation. This polynomial is solved to obtain a value for the scale factor ratio at a given redshift $\tilde{y}(z)$. Solving the equation numerically, we have to select the physical branch by ensuring that $0 < \tilde{y}(z) \leq 1$. Furthermore, we note that $\tilde{y} \rightarrow 1$ when the matter sources are vanishing ie. $z \rightarrow -1$.

We suppress the z dependence from now on writing $\tilde{y}(z) = \tilde{y}$. The effective dark energy component, which evolves with the scale ratio $\tilde{y}$, is given by:
\begin{align}
    \Omega_{\rm DE}(\tilde{y}) = \, \Omega_{\Lambda} - 3\, \sin^2(\theta) \, \Omega_{\rm FP}  \, \left(1 - \tilde{y} \right) \times
    \left(1 + A \left(1 - \tilde{y} \right) + \frac{B}{2} \left(1 - \tilde{y} \right)^2\right)\,.
\end{align}
Additionally, the effective mass of the spin-2 mode on the FRW background is expressed as:
\begin{align}
   \frac{m_{\rm FP}^2(\tilde{y})}{H_0^2} =  3 \, \Omega_{\rm FP} \,  \left( (1 + 2 A + B) \tilde{y} - 2(A + B) \tilde{y}^2 + B \tilde{y}^3 \right) \times  \left(\frac{1}{\tilde{y}^2} + \left(1 - \frac{1}{\tilde{y}^2} \right) \sin^2(\theta) \right)\,.
\end{align}
Finally, the effective graviton mixing parameter that evolves with y is given by: 
\begin{align}
   \sin^2(\theta)(\tilde{y}) = \frac{ \tilde{y}^2}{\left(\tilde{y}^2 -1 + \frac{1}{\sin^{2}(\theta)}  \right)}\,.
\end{align}

This parametrisation is convenient for the numerical analysis of the dark energy EOS evolution and is used for the data fitting procedures described in Sec.~\ref{sec:fittoobservations}. In particular, they allow for a straightforward implementation of the constraints from the Vainshtein mechanism, ensuring that the local modifications of gravity remain suppressed while enabling the time evolution of the dark energy component in the cosmological background. See Ref.~\cite{Hogas:2021fmr} for the allowed parameter values of $A$ and $B$. Assuming a spacially flat universe, we can eliminate $\Omega_{r0}$ by use of the Friedman equation. At late times, when $\Omega_{r0}$ is subdominant, we can neglect it and eliminate $\Omega_{m0}$ by use of Eq.~\ref{eq:FR1} instead.
Now given the above equations and a set of parameters ($\Omega_{m0}$), $\Omega_\Lambda$, $\Omega_{\rm FP}$, $\theta$, $A$, and $B$,
we can describe the evolution of the dark energy EOS state by:
\begin{align}
   w_{\rm DE}\left[z\right] =  -1 -  \frac{ \tilde{y}^2}{\left(\tilde{y}^2 -1 + \frac{1}{\sin^{2}(\theta)}  \right)} \, \frac{ \Omega_{\mathrm{source}}\left[z\right]}{ {\Omega_{\mathrm{DE}}(\tilde{y})}} \left(1 - \frac{2 H_0^2 \left( \Omega_{\mathrm{source}}\left[z\right]+ \Omega_{\mathrm{DE}}(\tilde{y}) \right)}{ {m}_{\mathrm{FP}}^2(\tilde{y})}\right)^{-1}\,.
\end{align}
%

\bibliography{literature}
\end{document}